\documentclass[english,12pt,draftcls,onecolumn]{IEEEtran}
\usepackage[T1]{fontenc}
\usepackage[latin9]{inputenc}
\usepackage{verbatim}
\usepackage{amsthm}
\usepackage{amsmath}
\usepackage{amssymb}
\usepackage{graphicx}
\usepackage{color}
\usepackage{cite}
\makeatletter

\IEEEaftertitletext{\vspace{-2\baselineskip}}

  \theoremstyle{plain}
  \newtheorem{prop}{\protect\propositionname}

 \newtheorem{lem}{Lemma}
 \newtheorem{thm}{Theorem}
  \newtheorem{defn}{Definition}

\makeatother

\makeatother

\usepackage{babel}
  \providecommand{\propositionname}{Proposition}

\begin{document}

\title{Energy-Neutral Source-Channel Coding\\
with Battery and Memory Size Constraints
\thanks{Paolo Castiglione is with AKG Acoustics GmbH, Vienna, Austria. Gerald Matz is with Vienna University of Technology, Vienna, Austria. Funding by FWF projects S10606 and S10607. The work of P. Castiglione was also supported by FTW Forschungszentrum Telekommunikation Wien project I0. The authors thank Prof.~Osvaldo Simeone for his valuable help in improving the quality of this work.}}

\author{Paolo Castiglione and Gerald Matz}

\maketitle
\begin{abstract}

We study energy management policies for the compression and transmission of source data
collected by an energy-harvesting sensor node with a finite energy buffer (e.g., rechargeable battery) and
a finite data buffer (memory) between source encoder and channel encoder.
The sensor node can adapt the source and channel coding rates depending on the observation and channel states.
In such a system, the absence of precise information about the amount of energy available in the future is a key challenge.
We provide analytical bounds and scaling laws for the average distortion that depend on the size of
the energy and data buffers. We furthermore design a resource allocation policy that achieves almost optimal distortion scaling.
Our results demonstrate that the energy leakage of state of art energy management policies can be avoided by
jointly controlling the source and channel coding rates.
\end{abstract}

\section{Introduction}

Energy harvesting techniques \cite{chalasani} enable the design of completely autonomous wireless sensor networks (WSN).
However, fluctuations in the amount of the energy being harvested call for resource management policies that 
achieve a trade-off between short-term metrics like delay and data queue length 
and long-term performance indicators like throughput and average distortion (see \cite{Keo12} and references therein).

In a WSN, an additional challenge is the fact that the energy consumption of source compression is in the same order 
as that of transmission. Even without energy harvesting this allows for energy savings via a joint energy management 
for source coding and transmission \cite{HE06,Barr,LU03}.
These results have been extended to fluctuating energy sources in \cite{Castiglione,tappa}.

In this paper, we consider a single sensor node and adopt the model from \cite{Castiglione}, where
an \textit{energy buffer}, e.g., a rechargeable battery, stores the harvested energy.
In each time slot, the node acquires and compresses an observation with a suitably adapted rate. 
The observation is characterized by a time-varying state, e.g., observation signal-to-noise ratio (SNR). 
Then, the node stores the source coder output bits in a \textit{data buffer} (memory).
Furthermore, it transmits to the destination a certain number of bits from the data buffer, using
a suitably adapted channel coding rate. The transmission channel is characterized by an instantaneous channel SNR. 

In previous work \cite{Castiglione}, we characterized optimal energy management policies that achieve minimum distortion for 
the extreme cases where the energy and data buffer are either infinite or very small. For infinite buffer size, where
the stability of the data queue needs to be guaranteed, the optimal policies independently allocate energy to the source and channel encoders.
On the other hand, for the case of very small buffer size, a joint energy allocation by means of dynamic programming was found to be optimal.

In this paper, we consider finite buffers and use large deviation
tools for our analysis that were developed in the seminal work of Tse \cite{Tse}. 
Compared to dynamic programming, these tools have the advantage of not
suffering from the curse of dimensionality.
In \cite{Tse}, only the source coding is taken into account in the sensor policy,
which thus amounts to choosing a point on the rate-distortion curve. 
Neither the problem of maintaining energy-neutrality, nor fluctuations of the available energy, nor 
optimal resource allocation among source and channel encoder have been addressed in \cite{Tse}.

In this work, we claim that distortion optimality can be achieved via a joint energy management for the source and channel encoders. In particular, we provide analytical bounds on the average distortion achievable with an energy-harvesting sensor, and on the scaling laws of the achievable average distortion with respect to buffer size. 
We further propose a joint energy management for source and channel encoding that 
asymptotically achieves the distortion lower bound and scales almost optimally with buffer size.
We emphasize that in related work \cite{Laneman,Koksal} on this topic a joint adaptation of the source code and of the channel code has not been considered since the bit stream entering the data buffer was modeled as exogenous (i.e., uncontrollable).



Other recent contributions \cite{tappa,huang} for multi-hop systems have shown that a good trade-off
between performance and buffer sizes can be found by using Lyapunov optimization techniques
that do not require knowledge of the statistics of the system states. In particular, \cite{tappa}
addresses the problem of jointly controlling distributed source coding and data transmission
and develops policies that achieve a distortion optimality gap that is inversely proportional 
to buffer size. In contrast to our work, the optimality of such policies is not discussed in \cite{tappa}. 

\vspace{-2mm}
\section{System Model\label{sec:System-Model}}

\begin{figure}[t]
\centering{\includegraphics[clip,width=8.5cm]{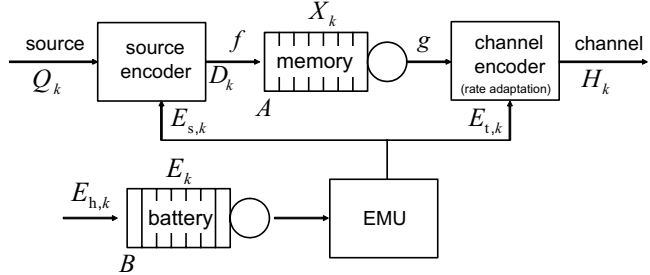}}

\vspace{-1mm}
\caption{Model for the energy harvesting sensor.}

\vspace{-4mm}
\label{Flo:model}
\end{figure}

We consider a system in which a single energy-harvesting sensor node communicates with a
single receiver. A block diagram of the sensor node is depicted in Fig.\,\ref{Flo:model}.
It essentially consists of a source encoder, a transmitter, an energy buffer, a data buffer,
and an energy management unit (EMU).

\vspace*{1mm}
{\it Energy buffer.} In our model, the sensor operation is structured in time slots (indexed by $k$). 
The energy harvested in slot $k$, denoted $E_{\mathrm{h},k}\in\mathbb{R}_{+}$, 
is accumulated in an energy buffer of finite size $B$, hence-forth also referred to as battery. 
For convenience, all energies are normalized by the number $N$ of channel uses per slot
(i.e., the number of symbols transmitted per time slot).
The harvested energy $E_{\mathrm{h},k}$ is assumed to be a discrete stationary
irreducible aperiodic Markov process. The steady-state probability density function (pdf) of $E_{\mathrm{h},k}$
is denoted $p_{E}(e)$. The energy $E_{k+1}$ available in the battery
for use in slot $k+1$ evolves as
\begin{equation}
E_{k+1}=\min\!\big\{B,\left[E_{k}-\left(E_{\mathrm{s},k}+E_{\mathrm{t},k}\right)\right]^{+}+E_{\mathrm{h},k}\big\},\label{eq:coda1}
\end{equation}
where $[x]^{+} = \max \{0,x\}$. Here, 
$\left[E_{k}-\left(E_{\mathrm{s},k}+E_{\mathrm{t},k}\right)\right]^{+}$
is the residual energy from the previous slot,
with $E_{\mathrm{s},k}$ and $E_{\mathrm{t},k}$ denoting the energies allocated in slot $k$ for 
source encoding and transmission, respectively.
We do not take into account the energy consumed by channel encoding and channel state acquisition, 
since they are typically small compared to the transmit energy in the scenario considered.
Energy-neutrality amounts to the constraint $E_{\mathrm{s},k}+E_{\mathrm{t},k}\leq E_{k}$. 

\vspace*{1mm}
{\it Source encoder.} The sensor takes $M$ measurements per time slot. The quality of these measurements is characterized by a parameter sequence $Q_{k}\in\mathcal{Q}$, which is assumed to form a discrete stationary irreducible aperiodic Markov process. 
As an example, $Q_{k}$ could model the measurement SNR, which may change over time due to source movement or environmental factors.
The set $\mathcal{Q}$ is assumed to be discrete and finite. The steady-state probability mass function (pmf) for $Q_{k}$ is 
denoted $\Pr(q)=\Pr(Q_{k}=q)$, $q\in\mathcal{Q}$. 
Due to sampling, analog-to-digital conversion, and compression
the sensor acquires the source in a lossy fashion.
The loss is captured by the distortion $D_{k}\in\mathbb{R}^{+}$, 
obtained from a given distortion metric such as the mean square error (MSE).
The bit stream resulting at the source encoder output is stored in a data buffer, subsequently also referred to as memory.

The number of bits produced by the source encoder within slot $k$ is given by 
$f^{(q)}(D_{k},E_{\mathrm{s},k})=f(D_{k},E_{\mathrm{s},k},Q_{k}=q)$. Here,
the rate-distortion-energy function $f$ models the dependence of the source encoder output on the distortion level $D_{k}$, the
allocated energy $E_{\mathrm{s},k}$, and the observation state $Q_{k}$. 
The function $f^{(q)}(D_{k},E_{\mathrm{s},k})$ is assumed (for any $q\in\mathcal{Q}$) to be continuous, differentiable, and separately strictly convex and non-increasing in $D_{k}$ and $E_{\mathrm{s},k}$. Example
rate-distortion-energy functions 
are provided in \cite{Castiglione}. 
Conventional rate-distortion functions are special cases without dependence on $E_{\mathrm{s},k}$.

\vspace*{1mm}
{\it Transmitter.} The channel between sensor and destination is characterized by a a discrete stationary irreducible aperiodic Markov process $H_{k}\in\mathcal{H}$ that changes slowly over time (e.g., block-fading). The pmf of $H_{k}$ is given by $\Pr(h)=\Pr(H_{k}=h)$, $h\in\mathcal{H}$. The transmitter uses the channel $N$ times per slot. A maximum number 
$g^{(h)}(E_{\mathrm{t},k})=g\left(H_{k}=h,E_{\mathrm{t},k}\right)$ of bits per slot can be communicated successfully to the destination. For any $h\in\mathcal{H}$, the channel rate function $g^{(h)}(E_{\mathrm{t},k})$ is assumed to be continuous, differentiable, strictly concave, and non-decreasing in $E_{\mathrm{t},k}$; furthermore, $g^{(h)}(0)=0$.
We consider rate-adaptive transmission schemes that achieve arbitrarily small block error probabilities. 
An example for $g$ is given by the Shannon capacity of the additive white Gaussian noise (AWGN) channel with SNR $h$. 
However, the channel-rate function can also model the rate of channel codes with a non-zero gap to Shannon capacity.
The number of bits actually transmitted using the allocated energy $E_{\mathrm{t},k}$ is given by
$\min\!\big\{X_{k},g^{(h)}(E_{\mathrm{t},k})\big\}$. 

\vspace*{1mm}
{\it Data buffer.} 
The size and queue length of the data buffer are denoted by $A$ and $X_{k}$, respectively.
The data queue length evolves as
\begin{equation}
X_{k+1}=\min\!\big\{A,\left[X_{k}-g(H_{k},E_{\mathrm{t},k})\right]^{+}+f(D_{k},E_{\mathrm{s},k},Q_{k})\big\}.
\label{eq:coda2}
\end{equation}
The source encoder increases the data queue length by $f(D_{k},E_{\mathrm{s},k},Q_{k})$ bits while 
the transmitter decreases the queue length by transmitting $g(H_{k},E_{\mathrm{t},k})$ bits.
When all parameters except $E_{\mathrm{s},k}$ and $E_{\mathrm{t},k}$ are fixed, \eqref{eq:coda2} captures the trade-off that results from splitting the available energy between the source encoder and the transmitter. 
Ideally, it is desirable to decrease $f(D_{k},E_{\mathrm{s},k},Q_{k})$ by increasing $E_{\mathrm{s},k}$
and simultaneously increase $g(H_{k},E_{\mathrm{t},k})$ by increasing $E_{\mathrm{r},k}$. 
However, due to energy neutrality $E_{\mathrm{s},k}$ and $E_{\mathrm{t},k}$ cannot be simultaneously increased without bounds.
 
 

\vspace{-2mm}
\section{Problem Statement and Main Results\label{sub:Policy Definition}}

\vspace{-2mm}
\subsection{Problem Statement}
\vspace{-1mm}

The results obtained in what follows are based on the assumption that the buffer sizes $A$ and $B$ are much larger than the maximum 
variation of the respective buffer states, i.e., ($A \gg \max |X_{k+1}-X_{k}|$ and $B \gg \max |E_{k+1}-E_{k}|$).
We further note that the Markov assumption for 
the energy harvesting, for the observation state, and for the channel state 
generalizes the memoryless assumption that was used in related work \cite{Laneman,Keo12}, and is ispired by recent models for real harvesting processes \cite{Ho10} as well as by well-established models for the wireless channel \cite{Sade08}. 
An extension to even more general models is beyond the scope of this paper. 

The EMU has to prescribe the distortion $D_{k}$ and the energies $E_{\mathrm{s},k}$ and $E_{\mathrm{t},k}$
to be allocated to the source encoder and the transmitter, respectively. It does so using the combined state
of the energy buffer, the data buffer, the source, and the channel, formally $S_{k}=\{E_{k},X_{k},Q_{k},H_{k}\}$.
More specifically, the EMU uses a policy $\pi=\left\{ \pi_{k}\right\} _{k\geq1}$
where $\pi_{k}=\left\{ D_{k}(S^{(k)}),E_{\mathrm{s},k}(S^{(k)}),E_{\mathrm{t},k}(S^{(k)})\right\} $
determines the parameters $(D_{k},E_{\mathrm{s},k},E_{\mathrm{t},k})$ in the $k$th time slot based on the
present and past states $S^{(k)}=\left\{ S_{1},\ldots,S_{k}\right\}$.


To decrease the distortion, the sensor can either use more compression energy,
thereby faster discharging the energy buffer, 
or compress less, thereby faster filling the data buffer 
(which necessitates an increase of the transmission energy to empty the buffer). 
Therefore, any policy $\pi$ amounts to a trade-off between the distortion performance and the
risk of an energy buffer drain or data buffer overflow. 
If the battery is empty or the memory is full,
a packet is lost and the maximum distortion $D_{\max}$ is accrued.
The long-term average distortion achieved with policy $\pi$ is defined as 
\begin{equation}
\bar{D}^{\pi} = \limsup_{n\rightarrow\infty}\frac{1}{n}\sum_{k=1}^{n} \mathbb{E}^{\pi}[D_{k}].
\label{eq:av_dist-1a}
\end{equation}
The optimal EMU policy $\pi_\text{opt}$ achieves the minimum distortion
$\bar{D}^{\min} = \min_{\pi} \bar{D}^{\pi}$; hence, $\bar{D}^{\pi}\ge \bar{D}^{\min}$ for all $\pi$.


\vspace{-1mm}
\subsection{Lower bound on the achievable distortion \label{sec:Stability-1}}

We define, conditional on the observation state $Q_{k}=q$, the long-term average source encoder energy 
$E_{\mathrm{s}}^{(q)}=\liminf_{n\rightarrow\infty}\frac{1}{n}\sum_{k=1}^{n}\mathbb{E}[E_{\mathrm{s},k} | Q_{k}=q]$ 
and the long-term average distortion
$D^{(q)}=\liminf_{n\rightarrow\infty}\frac{1}{n}\sum_{k=1}^{n}\mathbb{E}[D_{k}| Q_{k}=q]$.
Furthermore, the long-term average transmit energy conditional on the channel state $H_{k}=h$ is defined as
$E_{\mathrm{t}}^{(h)}=\liminf_{n\rightarrow\infty}\frac{1}{n}\sum_{k=1}^{n}\mathbb{E}[E_{\mathrm{t},k} | H_{k}=h]$. Our main results are based on the following convex optimization problem.

\begin{defn}
\label{def:policy_general}
\em The convex optimization problem $\text{CP}(\delta_\text{d},\delta_\text{e})$ is defined as follows: 
\allowdisplaybreaks
\begin{align}
& \underset{D^{(q)}\!, E_{\mathrm{s}}^{(q)}\!, E_{\mathrm{t}}^{(h)}\!,\hspace*{.15mm} \alpha}{\min}\;\sum_{q}\Pr(q)D^{(q)} \nonumber \\[1mm]
&\mathrm{subject\ to} \nonumber \\[1mm]
& \displaystyle
\sum_{q}\Pr(q)f^{(q)}\!\big(D^{(q)},E_{\mathrm{s}}^{(q)}\big)
- \sum_{h}\Pr(h)g^{(h)}\!\big(E_{\mathrm{t}}^{(h)}\big) \le \delta_\text{d},\label{constraint1}\\[0mm]
&\displaystyle
\sum_{q}\Pr(q)E_{\mathrm{s}}^{(q)}\leq\left(1\!-\!\alpha\right)\!\left(\mathbb{E}\left[E_{\mathrm{h},k}\right]+\delta_\text{e}\right)\!,\label{constraint2}\\
&\displaystyle
\sum_{h}\Pr(h)E_{\mathrm{t}}^{(h)}\leq\alpha\!\left(\mathbb{E}\left[E_{\mathrm{h},k}\right]+\delta_\text{e}\right)\!,\label{constraint3}\\
&\displaystyle
D^{(q)}\geq 0,\quad E_{\mathrm{s}}^{(q)}\geq 0,\quad E_{\mathrm{t}}^{(h)} \geq 0, \quad 0<\alpha<1.\nonumber
\end{align}
\end{defn}
Here, $\delta_\text{d}$ and $\delta_\text{e}$ can be interpreted as the
incremental and decremental drifts\footnote{The drift $\delta_\text{d}$ is the long-term expected difference between the size of the data buffer \emph{input}
and the size of the data buffer \emph{output}. Vice versa, the drift $\delta_\text{e}$ is the long-term expected difference between the size of the energy buffer \emph{output}
and the size of the energy buffer \emph{input}. For the definition of drift, the buffer is assumed to be unbounded. See Appendix for a more formal definition.} for the data buffer and for the energy buffer, respectively. The problem $\text{CP}(\delta_\text{d},\delta_\text{e})$ minimizes the inferior limit of the long-term expected distortion, thereby identifying the values of the above defined long-term expectations $D^{(q)}$, $E_{\mathrm{s}}^{(q)}$, $E_{\mathrm{t}}^{(h)}$  and of the associated parameter $\alpha$.  The latter parameter denotes the ratio between the long-term expected energy spent for transmission and the long-term overall expected energy spent for transmission and source coding. For the problem $\text{CP}(\delta_\text{d},\delta_\text{e})$ with $\delta_\text{d}\leq0$ and $\delta_\text{e}\leq0$, condition (\ref{constraint1}) is necessary for the \textit{mean rate stability} of the data queue \cite{neely2}, and conditions (\ref{constraint2})-(\ref{constraint3}) are necessary to meet the energy neutrality requirement. We note that the problems $\text{CP}(\delta_\text{d},\delta_\text{e})$ with $\delta_\text{d}>0$ and $\delta_\text{e}>0$ can be viewed as relaxations of $\text{CP}(0,0)$. 

Using the results in \cite{Castiglione}, we establish a lower bound on the achievable long-term distortion. 
The proofs of this and the subsequent results are provided in the Appendix.

\begin{prop}
\label{pro:minimumdist}
Let $\bar{D}^{*}$ denote the minimum of the problem $\mathrm{CP}(0,0)$ (i.e., with zero drift, $\delta_\text{d}=0$, $\delta_\text{e}=0$).
Then, the minimum distortion 
is lower bounded as $\bar{D}^{\min} \geq\bar{D}^{*}$. 
\end{prop}

We note that it suffices to prove this result for the special case 
of infinite data and energy buffer, i.e., $A=\infty$ and $B=\infty$. 
This also establishes the bound for finite buffer sizes 
since the assumption $A<\infty$ and $B<\infty$ is more restrictive (an infinite buffer can always mimic a finite buffer) and hence cannot lead to a smaller achievable distortion. The proof substantially demonstrates by means of Jensen inequality that  condition (\ref{constraint1}) is necessary to meet the \textit{mean rate stability} of the infinite data queue \cite{neely2}. 


The next result provides a lower bound on the scaling behavior of the difference $\bar{D}^{\min}-\bar{D}^{*}$, showing how $\bar{D}^{\min}$ converges to $\bar{D}^{*}$ when buffer size increases.%
\footnote{We use the following notation to compare the growth of two sequences
$a_{n}$ and $b_{n}$ as $n$ increases: $a_{n}=O(b_{n})$ if $a_{n}/b_{n}<c$  for large enough $n$ and
some constant $c$; $a_{n}=\Omega(b_{n})$ if $b_{n}=O(a_{n})$.}
\begin{prop}
\label{pro:scaling_law}
For any EMU policy $\pi$
there is 
\[
\bar{D}^{\min}-\bar{D}^{*} = \, \Omega\big(A^{-2}\big)+\Omega\big(B^{-2}\big) .
\]
\end{prop}
This results states that the optimality gap $\bar{D}^{\min}-\bar{D}^{*}$ is asymptotically bounded below by $c_1A^{-2}+c_2B^{-2}$ (here, $c_1$ and $c_2$ are constants). Thus, with increasing buffer size, $\bar{D}^{\min}$ cannot converge to the minimum distortion $\bar{D}^{*}$ at a rate faster than $c_1A^{-2}+c_2 B^{-2}$. This result is not intuitive. The key idea behind the proof provided in the Appendix entails the manipulation of appropriate balance equations for each buffer as done in \cite[2.4.1]{Tse}.

\vspace{-1mm}

\subsection{Distortion achievable with finite buffer size\label{sub:finite sizes}}


We now present a stationary EMU policy $\pi^\text{o} =\left\{ \pi_{k}^\text{o}\right\} _{k\geq1}$
that only depends (in a deterministic manner) on the current state $S_k$ 
and performs close to the lower bound established in Proposition 2.
This policy enforces drifts depending on hyper states that indicate whether the queues are more or less than half full. These hyper states are captured by the indices $n={\rm I}\{X_{k}\geq A/2\}$ and
$m={\rm I}\{E_{k} < B/2\}$.\footnote{The indicator function ${\rm I}\{\cdot\}$ equals $1$ if the argument is true and $0$ otherwise.}
The data queue drift and the energy buffer drift then equal $\delta_\text{d}=(-1)^{n}\frac{\beta_{1}\ln A}{A}$
and $\delta_\text{e}=(-1)^{m}\frac{\beta_{2}\ln B}{B}$, respectively, with $\beta_{1}$ and $\beta_{2}$ 
 sufficiently large constants (see Appendix). 
The sign of these drifts ensures that the buffer states are pushed towards the respective center levels $A/2$ and $B/2$. 
Furthermore, the drift magnitude depends on the size of the respective buffer.
For example, the data queue drift decreases with increasing buffer size $A$.
This is intuitive since smaller buffers tend to become full faster and hence require a stronger drift
to avoid overflow. 
The same reasoning applies to the battery drift.

\begin{figure}[t]
\centering{\includegraphics[clip,width=13cm]{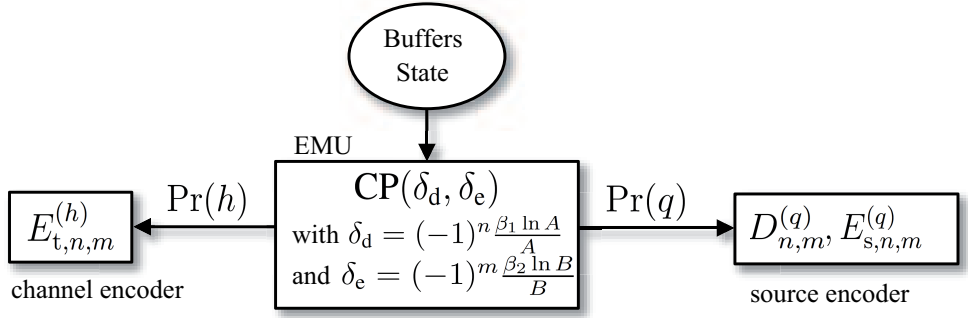}}

\vspace{-1mm}
\caption{Illustration of the EMU policy (\ref{eq:policy_optimal_scaling}).}

\vspace{-4mm}
\label{Flo:policy}
\end{figure}

\begin{defn}
\label{def:policy_optimal_scaling}
\em We define the policy $\pi_{k}^\text{o}=\{D_{k},E_{\mathrm{s},k},E_{\mathrm{t},k}\}$ by
\begin{align}
D_{k}& =D_{n,m}^{(q)}, \,\,\,\mathrm{for}\,\,\,Q_k=q, \nonumber \\
E_{\mathrm{s},k} & =\min\!\big\{ (1\!-\!\alpha_{n,m} )E_{k},E_{\mathrm{s},n,m}^{(q)}\big\}, \,\,\,\mathrm{for}\,\,\, Q_k=q,
\label{eq:policy_optimal_scaling}
\\
E_{\mathrm{t},k} & =\min\!\big\{\alpha_{n,m}  E_{k},E_{\mathrm{t},n,m}^{(h)}\big\}, \,\,\,\mathrm{for}\,\,\, H_k=h,\nonumber
\end{align}
where the parameters $D_{n,m}^{(q)}$, $E_{\mathrm{s},n,m}^{(q)}$, $E_{\mathrm{t},n,m}^{(h)}$, and $\alpha_{n,m}$ are obtained by solving the 
optimization problem $\text{CP}(\delta_\text{d},\delta_\text{e})$ from Definition 1 
with the drifts chosen as
$\delta_\text{d} = (-1)^{n}\frac{\beta_{1}\ln A}{A}$
and
$\delta_\text{e} = (-1)^{m}\frac{\beta_{2}\ln B}{B}$. 
\end{defn}

For given rate functions $f$ and $g$ and given statistics for $Q_k$ and $H_k$, the above problem must be solved for all four possible hyper states
using standard convex optimization tools and the resulting parameters of the policy can then be stored in a lookup table. As illustrated in Fig.\,\ref{Flo:policy}, the source code is determined by the distortion $D_{n,m}^{(q)}$ and the energy consumption $E_{\mathrm{s},n,m}^{(q)}$ depending on the state of the source $q$. On the other hand, the channel code is determined by the energy consumption $E_{\mathrm{t},n,m}^{(h)}$ depending on the channel state $h$. If the energy in the battery is not sufficient to provide $E_{\mathrm{s},n,m}^{(q)}$ and $E_{\mathrm{t},n,m}^{(h)}$, policy (\ref{eq:policy_optimal_scaling}) assigns the residual energy $E_{k}$ to the source encoder and to the channel encoder according to parameter $\alpha_{n,m} $. The next result assesses the performance of the EMU policy defined above.

\begin{prop}
\label{pro:achievable_policy}The policy $\pi^\text{o}$ 
achieves a long-term average distortion \textup{$\bar{D}^{\pi^\text{o}}$} that
approaches $\bar{D}^{*}$ as $O\!\left( A^{-2}\ln^2\! A \right)+O\!\left(B^{-2}\ln^2\! B\right)$.
\end{prop}

Proposition 3 states that the scaling behavior of the long-term average distortion achieved with 
the policy $\pi^\text{o}$ is almost optimal (cf.~Proposition 2).
More specifically, the optimality gap $\bar{D}^{\pi^\text{o}}-\bar{D}^{*}$
converges to zero as 
$O\left( A^{-2}\ln^2\! A \right)+O\left(B^{-2}\ln^2\! B\right)$. This scaling behaviour can be interpreted as the truncation of the Taylor representation of the optimality gap to the second order derivative with respect to the drifts in each buffer. The second order component dominates the performance because the first order component cancels out due to the fact that the drifts in each buffer have opposite signs, are equal in magnitude and occur with asymptotically equal probability (see Appendix).
Moreover, the data and energy buffer drifts imposed by $\pi^\text{o}$ keep the probabilities 
of battery depletion and memory overflow   
small (but different from zero), such that they become asymptotically negligible. More complex EMU policies, for instance, that can online adapt the source-channel code and the associated energies, might force these probabilities 
to zero. However, according to Proposition 2, any other policy, even if more complex and adaptive, cannot perform substantially better than 
$\pi^\text{o}$.

We additionally observe that the source and channel encoder parameters are jointly adapted over time. This is consistent with the dynamic programming solution in \cite{Castiglione} for small memory/battery sizes. It is also interesting to note that the source encoder and the transmitter are \textit{separately} controlled as long as the hyper state of the buffers remains the same, for instance, as long as the data queue length is less than $A/2$ and the available energy is larger than $B/2$. 
Hence, as $A,B \rightarrow \infty$, a separate energy management for source encoding and transmission
remains optimal, which is consistent with the results in \cite{Castiglione}.

\begin{figure}[t]
\centering{\includegraphics[clip,width=8.5cm]{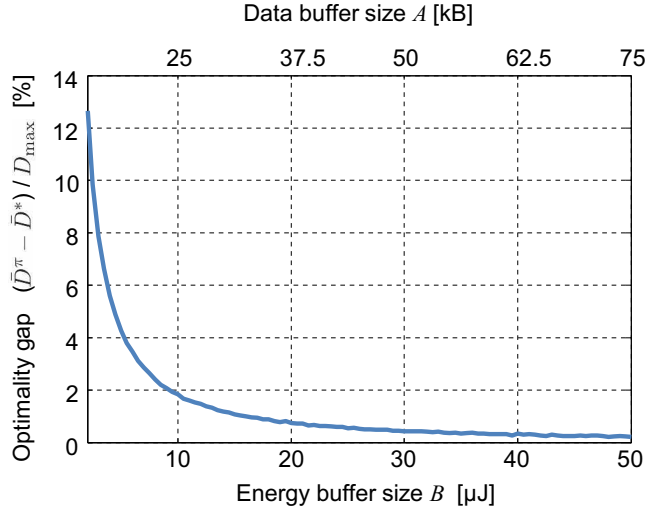}}
\vspace{-1mm}
\caption{Optimality gap versus data buffer size and energy buffer size.}

\vspace{-4mm}
\label{Flo:simu}
\end{figure}
\vspace{-2mm}
\section{Numerical Results}

We present Monte Carlo simulations in order to numerically assess the performance of the proposed EMU policy. 
We consider a system with slot duration $10\,$ms. In each, the sensor acquires $M=10^3$ noisy samples with SNR $Q_k$. The source encoder output bits are passed through a channel encoder and transmitted over an AWGN channel with $N=2\cdot 10^3$ channel uses (the transmission bandwidth thus is $200$kHz).

Using the model in \cite{HE06}, the source encoder is characterized by the rate-distortion-energy function 
$
f^{(q)}(D_{k},E_{\mathrm{s},k})=M\log_{2}\!\left( \frac{D_{\max}-D_{\mathrm{mmse}}}{D_k-D_{\mathrm{mmse}}}\right)\xi(E_{\mathrm{s},k}).
$ 
  Here, the first term is the information-theoretic rate-distortion limit for a zero-mean white Gaussian source with variance $D_{\max}$ 
and with minimum distortion (minimum MSE) $D_{\mathrm{mmse}}=\left(q+1/D_{\mathrm{max}}\right)^{-1}$. 
The function $\xi(E_{\mathrm{s},k})=\zeta\max\left\{1,\big(\frac{N}{M}\frac{E_{\mathrm{s},k}}{E_{\mathrm{s,max}}}\big)^{\eta}\right\}$
accounts for the rate increase 
incurred by practical (energy-limited) compression schemes.  
For our simulations we chose
$\zeta=2$, $\eta=-2/3$,  and a maximum energy consumption per slot of
$\frac{M}{N}E_{\mathrm{s,max}}=1\,\mathrm{nJ}=-7$dBm. 
The transmitter is characterized by the channel rate function
$g^{(h)}\left(E_{\mathrm{t},k}\right)=N\log(1+\frac{E_{\mathrm{t},k}}{h\Gamma\sigma})$,
where $H_{k}=h$ accounts for the path loss, $\Gamma=7$dB is the SNR gap to Shannon capacity of the 
(rate-adaptive) channel code, and $\sigma$ is the noise power (measured in J, here $-80\,$dBm). 

The source SNR $\mathcal{Q}=\{1,10\}$dB and the path loss $\mathcal{H}=\{40,50\}$dB are two-state Markov chains 
with uniform steady-state pmf and transitions from one state to the other happening with probability $0.1$.
The harvested energy $E_{\mathrm{h},k}$ is a uniformly distributed Markov chain with nine states, uniformly spaced in the interval $(0,100)\mu\mathrm{J/s}$ and with transition probability from each state to any of the other eight states equal to $0.05$. 

Fig.~\ref{Flo:simu} shows the optimality gap of the policy $\pi^\text{o}$ versus data and energy buffer size.
It is seen that the optimality gap can be decreased to 0.2\% of the source variance $D_{\mathrm{max}}$ by simultaneously increasing the energy buffer size and the data buffer size to realistic values
of 50$\mu$J and 75kB, respectively.\footnote{These are typical values for the capacitor and for the memory of a low-power sensor node.} Moreover, these results confirm the validity of the scaling behavior stated in Proposition 3.

\vspace{-1mm}
\section{Concluding Remarks}

In this paper we have proposed an energy management policy for energy-harvesting sensor nodes that achieves a close-to-optimal distortion scaling with respect to battery and memory size. Our large deviations results substantially differ from 
\cite{Laneman,Koksal}, which assumed the bits arriving in the data buffer to be exogenous (uncontrolled). In these papers, the average harvested energy was assumed to be strictly larger than
the average energy required to achieve the optimal utility. The energy buffer is therefore constantly filled, 
which implies that only the data buffer needs to be controlled, at the price that 
not all harvested energy is used. In contrast, our proposed energy management policy
jointly adapts the source code and the channel code, which leads to 
a separate control of the energy buffer and the data buffer and achieves the distortion lower bound without any detrimental energy leakage.

\section*{Appendix}
\subsection*{Proof of Proposition 1}

We prove this result for the special case of infinite data and energy buffer, i.e., 
$A = \infty$ and $B = \infty$. This also establishes the bound for finite buffer sizes since the assumption 
$A < \infty$ and $B < \infty$ is more restrictive (an infinite buffer can always mimic a finite buffer) and hence cannot lead to a smaller achievable distortion.


Without loss of generality we assume \textit{mean rate stability}. i.e., 
$\limsup_{n\rightarrow\infty}\frac{1}{n}\mathbb{E}[X_{k}]=0$
and $\mathbb{E}[X_{0}]<\infty$, both of which are always satisfied for the case 
of  finite data buffers.
Using \cite[Theorem 3]{neely2}, we now prove that mean rate stability implies the same necessary stability conditions 
as in \cite[Proposition 1]{Castiglione}. 

Since $f$ is convex non-increasing in $D_{k}$ and $E_{\mathrm{s},k}$ and $g$ is concave non-decreasing in $E_{\mathrm{t},k}$, 
Jensen's inequality implies 
\allowdisplaybreaks
\begin{align*}
\frac{1}{n}\sum_{k=1}^{n}\mathbb{E}[  f(D_{k},E_{\mathrm{s},k},Q_{k}) -g(H_{k},E_{\mathrm{t},k})]
 \geq
\sum_{q}\Pr(q) f^{(q)}(\bar{D}^{(q)}_n, \bar{E}^{(q)}_{\mathrm{s},n} ) -\sum_{h}\Pr(h) g^{(h)}( \bar{E}^{(h)}_{\mathrm{t},n}).
\end{align*}
with
\[
\bar{D}^{(q)}_n = \frac{1}{n}\sum_{k=1}^{n}\mathbb{E}[D_{k} | Q_{k}=q],\quad
\bar{E}^{(q)}_{\mathrm{s},n} = \frac{1}{n}\sum_{k=1}^{n}\mathbb{E}[E_{\mathrm{s},k} | Q_{k}=q],\quad  
\bar{E}^{(h)}_{\mathrm{t},n} = \frac{1}{n}\sum_{k=1}^{n}\mathbb{E}[E_{\mathrm{t},k} | H_{k}=h].
\]
Furthermore, we have
\begin{align*}
& \limsup_{n\rightarrow\infty} 
\sum_{q}\Pr(q) f^{(q)}(\bar{D}^{(q)}_n, \bar{E}^{(q)}_{\mathrm{s},n} ) = \sum_{q}\Pr(q) f^{(q)}( {D}^{(q)}, {E}^{(q)}_{\mathrm{s}} ),\\
& \limsup_{n\rightarrow\infty} -\sum_{h}\Pr(h) g^{(h)}(\bar{E}^{(h)}_{\mathrm{t},n}) = - \sum_{h}\Pr(h) g^{(h)}({E}^{(h)}_{\mathrm{t}}) ,
\end{align*}
with
$D^{(q)}=\liminf_{n\rightarrow\infty} \bar{D}^{(q)}_n$,
$E_{\mathrm{s}}^{(q)}=\liminf_{n\rightarrow\infty} \bar{E}^{(q)}_{\mathrm{s},n}$,
and
$E_{\mathrm{t}}^{(h)}=\liminf_{n\rightarrow\infty} \bar{E}^{(h)}_{\mathrm{t},n}$.
Using the fact that mean rate stability and \cite[(12)]{neely2}
imply $\limsup_{n\rightarrow\infty}\frac{1}{n}\sum_{k=1}^{n}\mathbb{E}[f(D_{k},E_{\mathrm{s},k},Q_{k})-g(H_{k},E_{\mathrm{t},k})]\leq0$,
we finally arrive at
\[
\sum_{q}\Pr(q)f^{(q)} (D^{(q)},E_{\mathrm{s}}^{(q)} ) \leq \sum_{h}\Pr(h)g^{(h)}(E_{\mathrm{t}}^{(h)}),
\]
which is the same necessary stability condition as in \cite[Proposition 1]{Castiglione}.
The proof of Proposition 1 thus follows from \cite[Proposition 1]{Castiglione}.

\subsection*{Proof of Proposition 2}

We first define some quantities that are instrumental for the proofs of Propositions 2 and 3.
The probabilities that a policy $\pi$ results in an empty energy buffer or a full data queue are respectively defined as
\begin{align}\label{eq:peb}
p^\pi_{\mathrm{EB}} & = \limsup_{n\rightarrow\infty}\frac{1}{n}\sum_{k=1}^{n}\Pr(\left[E_{k}- E_{\mathrm{s},k} - E_{\mathrm{t},k}\right]^{+}=0),\\
p^\pi_{\mathrm{FQ}} & = \limsup_{n\rightarrow\infty}\frac{1}{n}\sum_{k=1}^{n}\Pr(X_{k}=A)
\label{eq:pfq}
\end{align}
The \emph{decremental} drift $\delta_{\mathrm{e}}$ of the unbounded energy buffer queue process $\tilde{E}_{k+1}=\tilde{E}_{k}-E_{\mathrm{s},k}-E_{\mathrm{t},k}+E_{\mathrm{h},k}$, and the \emph{incremental} drift $\delta_{\mathrm{d}}$ of the unbounded data queue process $\tilde{X}_{k+1}=\tilde{X}_{k}+f(D_{k},E_{\mathrm{s},k},Q_{k})-g(H_{k},E_{\mathrm{t},k})$  are respectively defined as
\begin{flalign*}
\delta_{\mathrm{e}} & = \lim_{n\rightarrow\infty}\frac{1}{n}\sum_{k=1}^{n}\mathbb{E}[\tilde{E}_{k}-\tilde{E}_{k+1}],\\
\delta_{\mathrm{d}} & = \lim_{n\rightarrow\infty}\frac{1}{n}\sum_{k=1}^{n}\mathbb{E}[\tilde{X}_{k+1}-\tilde{X}_{k}].
\end{flalign*}
The drift $\delta_\text{d}$ can be viewed as the long-term expected difference between the size of the data buffer \emph{input}
and the size of the data buffer \emph{output}, whereas the drift $\delta_\text{e}$ can be viewed as the long-term expected difference between the size of the energy buffer \emph{output}
and the size of the energy buffer \emph{input}.

We denote the event of \emph{normal operation} by
$\mathcal{E}_k = \{ X_{k}<A\text{ and }\left[E_{k}-\left(E_{\mathrm{s},k}+E_{\mathrm{t},k}\right)\right]^{+}>0\} $
(i.e., the data buffer is not full and the energy buffer is not empty). The complementary event is denoted as
$\overline{\mathcal{E}}_k$ (i.e., either the data buffer is full or the energy buffer is empty). 
The instantaneous expected distortion can now be written as 
$\mathbb{E}^{\pi}\!\left[D_{k}\right]=\mathbb{E}^{\pi}\!\left[D_{k}|\mathcal{E}_k\right]\Pr(\mathcal{E}_k)
+\mathbb{E}^{\pi}\!\left[D_{k}|\overline{\mathcal{E}}_k\right]\Pr(\overline{\mathcal{E}}_k)$.
With $\Pr(\mathcal{E}_k)\le 1$, $\mathbb{E}^{\pi}\!\left[D_{k}|\overline{\mathcal{E}}_k\right]=D_{\max}$, and
the union bound $\Pr(\overline{\mathcal{E}}_k)\le \Pr(X_k=A) + \Pr(\left[E_{k}-\left(E_{\mathrm{s},k}+E_{\mathrm{t},k}\right)\right]^{+}=0)$,
we obtain the following upper bound on the long-term average distortion $\bar{D}^{\pi}$:
\begin{equation}
\bar{D}^{\pi}=\limsup_{n\rightarrow\infty}\frac{1}{n}\sum_{k=1}^{n}\mathbb{E}^{\pi}\!\left[D_{k}\right]
\,\;\leq\; D^{\mathrm{op}}+D_{\mathrm{max}}\left(p_{\mathrm{FQ}}^{\pi}+p_{\mathrm{EB}}^{\pi}\right).
\label{eq:av_dist}
\end{equation}
Here, we have used \eqref{eq:peb}, \eqref{eq:pfq}, and the long-term average distortion during normal operation,
\begin{equation}\label{eq:Dop_def}
D^{\mathrm{op}}=\limsup_{n\rightarrow\infty}\frac{1}{n}\sum_{k=1}^{n}
\mathbb{E}^{\pi}\!\left[D_{k}|\mathcal{E}_k\right].
\end{equation}
Note that here we assume that if the energy buffer is empty or the data buffer is full, a packet is lost and the maximum distortion 
$D_{\mathrm{max}}$ is accrued (i.e., the decoder treats this missing packet as an arbitrary vector, e.g., the mean of the source distribution). 
An energy buffer discharge or data buffer overflow could be handled in a more sophisticated manner, but this does not bear on our asymptotic analysis (see also \cite{Tse} for a similar reasoning).
The bound \eqref{eq:av_dist} indicates that studying the optimal convergence of $\bar{D}^{\pi}$ to the lower bound 
$\bar{D}^{*}$ (see Proposition 1) is equivalent to finding the optimal scaling laws for
i) the probabilities $p_{\mathrm{EB}}^{\pi}$ and $p_{\mathrm{FQ}}^{\pi}$ approaching zero and 
ii) the operational distortion $D^{\mathrm{op}}$ approaching $\bar{D}^{*}$.

In order to prove Proposition 2, we need the following two lemmas\footnote{The following notations are used to compare two sequences 
$a_{n}$ and $b_{n}$ as $n$ grows: $a_{n}=O(b_{n})$ if $a_{n}/b_{n}<c$ for all $n$ and
some constant $c$; $a_{n}=\Omega(b_{n})$ if $b_{n}=O(a_{n})$; $a_{n}=\Theta(b_{n})$
if $a_{n}=O(b_{n})$ and $b_{n}=O(a_{n})$; $a_{n}=o(b_{n})$ if $\lim_{n\rightarrow\infty}a_{n}/b_{n}=0$.}.

\begin{lem}
\label{lemma1}Let $B=\infty$ and consider an arbitrary control scheme
$\pi_{k}=\left\{ D_{k},E_{\mathrm{s},k},E_{\mathrm{t},k}\right\} $ that achieves $p_{\mathrm{FQ}}^{\pi}=o(1/A^{2})$.
Then, $\bar{D}^{\pi}-\bar{D}^{*}=\Omega(1/A^{2})$.
\end{lem}
Lemma \ref{lemma1} states that for an infinitely large energy buffer no control scheme can 
make both $p_{\mathrm{FQ}}^{\pi}$ and $\bar{D}^{\pi}$ converge at a rate faster 
than $1/A^{2}$. The proof of Lemma \ref{lemma1} is based on \cite[Proposition 2.4.1]{Tse}. 
Let us consider the optimization problem $\text{CP}(\delta_\mathrm{d},0)$, i.e., the problem formulated in Def.~1 with the
specific choice $\delta_\mathrm{e}=0$. Denote the minimum of $\text{CP}(\delta_\mathrm{d},0)$ by $D_{\mathrm{T}}(\delta_\mathrm{d})$.
 
According to Proposition 1, the solution to this problem with $\delta_{\mathrm{d}}=0$
equals the lower bound on the minimum achievable distortion. Notice that,
by the convexity \cite{Boyd} of the problem, 
$D_{\mathrm{T}}(\delta_\mathrm{d})$ 
is convex and non-increasing in $\delta_{\mathrm{d}}$. Moreover, using the same arguments
as in Proposition 1, it can be proved that there exists no policy $\pi$ that is able to achieve a long term
average distortion smaller than  $D_{\mathrm{T}}(\delta_\mathrm{d})$.
with a data queue drift smaller or equal to $\delta_{\mathrm{d}}$. 
Thus,  $D_{\mathrm{T}}(\delta_\mathrm{d})$ can be viewed as the lower limit of the distortion-drift region. 
This observation allows us to directly apply the proof of \cite[Proposition 2.4.1]{Tse} to Lemma \ref{lemma1}.

\begin{lem}
\label{lemma2}Let $A=\infty$, and consider an arbitrary control scheme
$\pi_{k}=\left\{ D_{k},E_{\mathrm{s},k},E_{\mathrm{t},k}\right\} $ that achieves $p_{\mathrm{EB}}^{\pi}=o(1/B^{2})$.
Then, $\bar{D}^{\pi}-\bar{D}^{*}=\Omega(1/B^{2})$.
\end{lem}
Lemma \ref{lemma2} is the counterpart of Lemma \ref{lemma1};
it states that, for an infinitely large data buffer size, no control
scheme can achieve a convergence rate faster than $1/B^{2}$ for both
$p_{\mathrm{EB}}^{\pi}$ and $\bar{D}^{\pi}$. The proof parallels that of Lemma \ref{lemma1}.

We now prove Proposition 2 by contradiction. Assume
that there exists a policy with $p_{\mathrm{EB}}^{\pi}=o(1/B^{2})$
and $p_{\mathrm{FQ}}^{\pi}=o(1/A^{2})$ that achieves $\bar{D}^{\pi}-\bar{D}^{*}=O(1/A^{2})+O(1/B^{2})$,
i.e., $\bar{D}^{\pi}-\bar{D}^{*}$ is asymptotically bounded \textit{above}
by $c_1/A^{2}+c_2/B^{2}$
(where $c_1$ and $c_2$ are constant factors).
Such a policy would violate Lemma \ref{lemma1} (or  Lemma \ref{lemma2})
as $B$ (or $A$) tends to infinity and hence cannot exist. This implies that
$\bar{D}^{\pi}-\bar{D}^{*}=\Omega(1/A^{2})+\Omega(1/B^{2})$,
which concludes the proof of Proposition 2.

\subsection*{Proof of Proposition 3}

In order to prove Proposition 3, we first need to recall some known results. 
Let us define the random walk $Z_{k+1}=Z_{k}+W_{k}$, $k\ge 0$, 
with $Z_0=0$ and $W_{k}$ a stationary, irreducible, and aperiodic Markov chain with states $w_{i}$, $1 \leq i\leq I$. 
The transition probability from state $w_{i}$ to state $w_{j}$ is denoted $p_{i,j}$ and 
$\Pr(w_{i})$ is the invariant distribution of $W_k$.
The drift $R=\mathbb{E}[W_{k}]$ of the random walk is assumed negative. 
The moment generating function of $W_{k}$ is given by
\[
\rho(r) = \mathbb{E}\left[\exp(r W_{k})\right]=\sum_{i=1}^{I} \Pr(w_{i}) \exp(r w_{i}).
\]
It can be shown that the function $\log\rho(r)$ 
(i.e., the cumulant-generating function)
has a unique positive zero at $r=r^{*}$.
\begin{thm}
(Wald's identity) 
Let K be the first
$k\geq1$ for which $Z_{k}\geq a\ge 0$ or $Z_{k}\leq b\le 0$.
Then 
\[
\mathbb{E}[\exp(r^{*}Z_{K})]=1,
\]
where $r^{*}$ is the unique positive root of
$\log\rho(r)$.
Furthermore,
\[
\mathbb{E}[K]\,\mathbb{E}[W_{k}]=\mathbb{E}[Z_{K}].
\]
\end{thm}
Using Theorem 1 we can compute the probability
$p=\Pr(Z_{K}\geq L)$ that a negative-drift random walk that starts at zero
will cross the barrier $L>0$ before returning  to the origin,
\[
p\mathbb{E}[\exp(r^{*}Z_{K})|Z_{K}\geq L]+(1-p)\mathbb{E}[\exp(r^{*}Z_{K})|Z_{K}\leq0]=1.
\]
Since $\mathbb{E}[\exp(r^{*}Z_{K})|Z_{K}\geq L]=\Theta( \exp(r^{*}L) )$
and $\mathbb{E}[\exp(r^{*}Z_{K})|Z_{K}\leq0]=\Theta(1)$ in the regime
of large $L$, we have 
\begin{gather*}
p=\Theta( \exp(-r^{*}L) ),
\qquad
\mathbb{E}[Z_{K}]=p\mathbb{E}[Z_{K}|Z_{K}\geq L]+(1-p)\mathbb{E}[Z_{K}|Z_{K}\leq0]=\Theta(1).
\end{gather*}
Hence, by Theorem 1, also the expected crossing time is dominated
by the return to the origin, i.e., $\mathbb{E}[K]=\mathbb{E}[Z_{K}]/\mathbb{E}[W_{k}]=\Theta(1)$.

Let us now define $D_{\text{T}}(\delta_\mathrm{d},\delta_\mathrm{e})$, 
as the minimum distortion in the convex problem $\text{CP}(\delta_\mathrm{d},\delta_\mathrm{e})$ (Def.~1).
According to Proposition 1, 
$\bar{D}^\star = D_{\mathrm{T}}(0,0)$ is a lower bound for
the minimum achievable distortion $D_{\text{T}}(\delta_\mathrm{d},\delta_\mathrm{e})$. 
The convexity of $\text{CP}(\delta_\mathrm{d},\delta_\mathrm{e})$ implies that $D_{\mathrm{T}}(\delta_\mathrm{d},\delta_\mathrm{e})$ is convex
and non-increasing in $(\delta_\mathrm{d},\delta_\mathrm{e})$. Moreover, it can be proved using the same arguments as in Proposition 1
that there exists no policy $\pi$ that is able to achieve a long-term average distortion $\bar{D}^{\pi}$ smaller than $D_{\mathrm{T}}(\delta_\mathrm{d},\delta_\mathrm{e})$ with a data queue drift and energy buffer drift less than or equal to $\delta_\mathrm{d}$ and $\delta_\mathrm{e}$, respectively.
Thus, $D_{\mathrm{T}}(\delta_\mathrm{d},\delta_\mathrm{e})$ can be viewed as a lower bound for the distortion-drift region.

Consider now the policy $\pi^\text{o}$ in Def.~2 and 
recall that the hyper states $\hat{X}_{k}=\mathrm{I}(X_k\ge A/2)\in\{ 0,1 \} $ and $\hat{E}_{k}=\mathrm{I}(E_k < B/2)\in\{ 0,1 \} $
indicate, respectively, whether the data buffer is more than half full and the energy buffer is more than half empty.
Both $\hat{X}_{k}$ and $\hat{E}_{k}$ can be shown to be irreducible and aperiodic Markov chains.
Furthermore, the 
data queue \emph{increment} process 
$W_{k}^{\mathrm{d}}(\hat{X}_{k})=f(D_{k},E_{\mathrm{s},k},Q_{k})-g(H_{k},E_{\mathrm{t},k})$
in each half of the data buffer is function of the aggregate Markov chain $(\hat{E}_{k},Q_{k},H_{k})$.
Hence, $W_{k}^{\mathrm{d}}(\hat{X}_{k})$ in each half of the data queue is is itself an irreducible and aperiodic Markov chain whose mean equals the drift $\delta_{\mathrm{d}}$. 
Similarly, the 
energy buffer \emph{decrement} process $W_{k}^{\mathrm{e}}(\hat{E}_{k})= E_{\mathrm{s},k} + E_{\mathrm{t},k}-E_{\mathrm{h},k} $ in each
half of the energy buffer is a function of the aggregate Markov chain $(\hat{X}_{k},E_{\mathrm{h},k},Q_{k},H_{k})$ 
and hence is itself an irreducible and aperiodic Markov chain whose mean equals the drift (as defined above), i.e., $\delta_{\mathrm{e}}$.
We next state a lemma that follows from \cite{Koksal} and relates the unique positive roots $r_{\mathrm{d}}^{*}(\hat{X}_{k})$
and $r_{\mathrm{e}}^{*}(\hat{E}_{k})$ of the cumulant-generating functions of 
$W_{k}^{\mathrm{d}}(\hat{X}_{k})$ and $W_{k}^{\mathrm{e}}(\hat{E}_{k})$, respectively.
\begin{lem}
\label{lemma3}Assume that the policy $\pi^\mathrm{o}$ is the unique optimal policy for the drifts $\delta_\mathrm{d}$ and $\delta_\mathrm{e}$.
Then
\begin{subequations}
\begin{align}\label{eq:diffdrift1}
\frac{dr_{\mathrm{d}}^{*}(\hat{X}_{k})}{d\delta_{\mathrm{d}}}\bigg|_{\delta_\mathrm{d}=\delta_\mathrm{e}=0}
= -\frac{2}{\mathrm{var}(W_{k}^{\mathrm{d}}(\hat{X}_{k})|\delta_\mathrm{d}=\delta_\mathrm{e}=0)},\\
\frac{dr_{\mathrm{e}}^{*}(\hat{E}_{k})}{d(\delta_{\mathrm{e}})}\bigg|_{\delta_\mathrm{d}=\delta_\mathrm{e}={0}}
= -\frac{2}{\mathrm{var}(W_{k}^{\mathrm{e}}(\hat{E}_{k})|\delta_\mathrm{d}=\delta_\mathrm{e}={0})}.
\label{eq:diffdrift2}
\end{align}
\end{subequations}
\end{lem}
The proof of Lemma \ref{lemma3} is based on the fact that 
the policy $\pi^\mathrm{o}$ depends smoothly on the drifts $(\delta_\mathrm{d},\delta_\mathrm{e})$ in the neighborhood of $\delta_\mathrm{d}=\delta_\mathrm{e}={0}$. This implies that $r_{\mathrm{x}}^{*}(\hat{X}_{k})$ and $r_{\mathrm{e}}^{*}(\hat{E}_{k})$
are smooth functions of the means of the respective increment/decrement processes, i.e., $\delta_\mathrm{d}$ and $\delta_\mathrm{e}$, too.
Since the supports of $W_{k}^{\mathrm{x}}(\hat{X}_{k})$ and $W_{k}^{\mathrm{e}}(\hat{E}_{k})$ are finite and
the policy $\pi^\mathrm{o}$ is almost always continuous and differentiable (like the waterfilling-like policy in \cite{Berry2}), 
the respective second-order derivatives around $\delta_\mathrm{d}=\delta_\mathrm{e}={0}$ almost always exist and are continuous.

To obtain $r_{\mathrm{d}}^{*}$ (we omit the argument $\hat{X}_{k}$ in what follows for notational simplicity),
we need to find the root of the cumulant generating function of 
$W_{k}^{\mathrm{d}}(\hat{X}_{k})$ with $\delta_{\mathrm{e}}=0$:
\[
\Lambda(r_{\mathrm{d}}^{*})=\log \rho (r_{\mathrm{d}}^{*}) =\log \left( \sum_{i=1}^{I} \Pr(w_{i}^{\mathrm{d}}) \exp(r_{\mathrm{d}}^{*}  w_{i}^{\mathrm{x}})\right)=0;
\]
here, $\Pr(w_{i}^{\mathrm{d}})$ is the invariant distribution of $W_{k}^{\mathrm{d}}(\hat{X}_{k})$,
which depends on $\delta_{\mathrm{d}}$. Denoting by 
$\kappa_n$
the $n$th cumulant of $W_{k}^{\mathrm{d}}(\hat{X}_{k})$
(i.e., the $n$th derivative of the cumulant generating function $\Lambda(r)$ 
at $r=0$), the Maclaurin expansion of $\Lambda(r_{\mathrm{d}}^{*})$ reads
\[
\Lambda(r_{\mathrm{d}}^{*})=\kappa_0 + \sum_{n=1}^{\infty} \kappa_n\frac { (r_{\mathrm{d}}^{*})^{n}} {n!}=\delta_{\mathrm{d}}r_{\mathrm{d}}^{*} + \sum_{n=2}^{\infty} \kappa_n\frac { (r_{\mathrm{d}}^{*})^{n}} {n!},
\]
where the second equality is obtained with $\kappa_0=0$ and $\kappa_1=\delta_{\mathrm{d}}$. 
Setting this expression equal to zero, dividing both sides by $r_{\mathrm{d}}^{*}$ and differentiating with respect to $\delta_{\mathrm{d}}$ yields
\[
\sum_{n=2}^{\infty} \kappa_n\frac { (n-1)(r_{\mathrm{d}}^{*})^{n-2}} {n!} \,\frac {dr_{\mathrm{d}}^{*}} {d\delta_{\mathrm{d}}} = -1.
\]
Since $r_{\mathrm{d}}^{*} \rightarrow 0$ as $\delta_{\mathrm{d}} \rightarrow 0$ (due to well-known properties of the moment generating function, see, e.g., \cite{Tse}), the above expression becomes
\[
\frac{\kappa_2}{2} \frac{dr_{\mathrm{d}}^{*}}{d\delta_{\mathrm{d}}}\bigg|_{\delta_\mathrm{d}=0}
= -1.
\]
By substituting $\kappa_2=\mathrm{var}(W_{k}^{\mathrm{d}}(\hat{X}_{k})|\delta_\mathrm{d}=\delta_\mathrm{e}=0)$,
we obtain \eqref{eq:diffdrift1} in Lemma \ref{lemma3}. The proof of 
\eqref{eq:diffdrift2} is analogous.

\def\Ku{K^\mathrm{u}}
Let us next consider the probability of a full data buffer $p_{\mathrm{FQ}}^{\pi^\mathrm{o}}$
obtained by adopting the policy $\pi^\mathrm{o}$. 
The times at which the data buffer becomes full can be viewed as the epochs of a renewal process \cite{BOR76}. 
The fullness probability can thus be written as $p_{\mathrm{FQ}}^{\pi^\mathrm{o}}=1/\mathbb{E}[Y]$, where $Y$ is the duration
between successive time instants at which the buffer becomes full. 
We define the random walk $Z_{k}^{\mathrm{u}}=Z_{k-1}^{\mathrm{u}}+W_{k}^{\mathrm{d}}(\hat{X}_{k}=1)$, $k=0,1,\dots$, 
in the upper half of the data buffer, with negative drift $\delta_\mathrm{d}$.
%
Let $\Ku$ be the smallest time index $k\geq1$ for which $Z_{k}^{\mathrm{u}}\geq 0$ 
or $Z_{k}^{\mathrm{u}}\leq-A/2$.
We then have (for similar arguments refer to \cite{Tse})
\begin{flalign}
\mathbb{E}[Y]= & \mathbb{E}[\Ku|Z_{0}^{\mathrm{u}}=0]  +\Pr\!\bigg[Z_{\Ku}^{\mathrm{u}}\leq-\frac{A}{2}\bigg|Z_{0}^{\mathrm{u}}=0\bigg]\nonumber \\
 & \cdot\int_{0}^{g^{\mathrm{max}}}\!\!\Pr\!\bigg[Z_{\Ku}^{\mathrm{u}}=-\frac{A}{2}-x\bigg|Z_{\Ku}^{\mathrm{u}}\leq-\frac{A}{2},Z_{0}^{\mathrm{u}}=0\bigg]\,\mathbb{E}\bigg[V\bigg|X_{0}=\frac{A}{2}-x\bigg]\,dx,\label{eq:2.3}
\end{flalign}
where $g^{\mathrm{max}}$ is the maximum undershoot relative to the center of the buffer (i.e., the maximum variation for the considered policy towards the empty buffer) and $V$ is the time it takes to fill up the buffer starting from an initial buffer state $X_{0}$.
We furthermore  define the time $J^{\mathrm{u}}$ to be the smallest $k\geq 1$ for which $Z_{k}^{\mathrm{u}}\leq0$
or $Z_{k}^{\mathrm{u}}\geq A/2$. Also, we define another random walk for the
dynamics in the lower half of the buffer (state $\hat{X}_k=0$),
$Z_{k}^{\mathrm{l}}=Z_{k-1}^{\mathrm{l}}+W_{k}^{\mathrm{d}}(\hat{X}_{k}=0)$, $k=0,1,\dots$,
with positive drift $\delta_{\mathrm{d}}$. 
The smallest $k\geq1$ such that $Z_{k}^{\mathrm{l}}\leq0$ or $Z_{k}^{\mathrm{l}}\geq A/2$ is denoted $K^{\mathrm{l}}$, and
the smallest $k\geq1$ such that $Z_{k}^{\mathrm{l}}\geq0$ or $Z_{k}^{\mathrm{l}}\leq-A/2$
is denoted $J^{\mathrm{l}}$. 
For $0\leq x<g^{\mathrm{max}}$ we have 
\allowdisplaybreaks
\begin{flalign}
\mathbb{E}\!\left[V\bigg|X_{0}=\frac{A}{2}-x\right]= & \mathbb{E}\!\left[P\bigg|X_{0}=\frac{A}{2}-x\right]
\nonumber \\ & 
+\int_{0}^{f^{\mathrm{max}}}\!\!\Pr\!\left[X_{P}=\frac{A}{2}+y\bigg|X_{0}\leq\frac{A}{2}-x\right]\mathbb{E}\!\left[V\bigg|X_{0}=\frac{A}{2}+y\right]dy,\label{eq:2.4}
\end{flalign}
where the integral is up to the maximum overshoot $f^{\mathrm{max}}$ 
relative to the center of the buffer (i.e. the maximum variation for
the considered policy towards the full buffer). The quantity $P$
is the first time instant for which the buffer becomes more than half full starting
from an initial buffer state $X_{0}$.

By conditioning on the event that the buffer becomes full or that the data queue leaves the upper buffer half,
we have for every $0\leq y<f^{\mathrm{max}}$
\allowdisplaybreaks
\begin{flalign}
\mathbb{E}\!\left[V\bigg|X_{0}=\frac{A}{2}+y\right]= & \mathbb{E}\!\left[J^{\mathrm{u}}\bigg|Z_{0}^{\mathrm{u}}=y\right] +\Pr\!\left[Z_{J^{\mathrm{u}}}^{\mathrm{u}}\leq0\bigg|Z_{0}^{\mathrm{u}}=y\right]\nonumber \\
 & \cdot\int_{0}^{g^{\mathrm{max}}}\Pr\!\left[Z_{J^{\mathrm{u}}}^{\mathrm{u}}=-w\bigg|Z_{J^{\mathrm{u}}}^{\mathrm{u}}\leq0\right]\mathbb{E}\!\left[V\bigg|X_{0}=\frac{A}{2}-w\right]dw.\label{eq:2.5}
\end{flalign}
From (\ref{eq:2.5}) it follows that for every $0\leq y<f^{\mathrm{max}}$ 
\allowdisplaybreaks
\begin{flalign*}
\mathbb{E}\!\left[V\bigg|X_{0}=\frac{A}{2}+y\right]
\leq \, 
\mathbb{E}\!\left[J^{\mathrm{u}}\bigg|Z_{0}^{\mathrm{u}}=y\right]
+\Pr\!\left[Z_{J^{\mathrm{u}}}^{\mathrm{u}}\leq0|Z_{0}^{\mathrm{u}}=y\right]\max_{0\leq w<g^{\mathrm{max}}}\mathbb{E}\!\left[V\bigg|X_{0}=\frac{A}{2}-w\right].
\end{flalign*}
Inserting this bound into \eqref{eq:2.4} yields for every $0\leq x<g^{\mathrm{max}}$ 
\allowdisplaybreaks
\begin{align*}
\mathbb{E}\!\left[V\bigg|X_{0}=\frac{A}{2}-x\right] 
\leq &\;\, \mathbb{E}\!\left[P\bigg|X_{0}=\frac{A}{2}-x\right]
+\max_{0\leq y<f^{\mathrm{max}}}\mathbb{E}\left[J^{\mathrm{u}}|Z_{0}^{\mathrm{u}}=y\right]\\
 & +\max_{0\leq y<f^{\mathrm{max}}}\Pr\!\left[Z_{J^{\mathrm{u}}}^{\mathrm{u}}\leq0|Z_{0}^{\mathrm{u}}=y\right]
 \max_{0\leq w<g^{\mathrm{max}}}\mathbb{E}\!\left[V\bigg|X_{0}=\frac{A}{2}-w\right].
\end{align*}
Taking the maximum with respect to $0\leq x<g^{\mathrm{max}}$
and noting that 
$1\,-\!\max\limits_{0\leq y<f^{\mathrm{max}}}\Pr\!\left[Z_{J^{\mathrm{u}}}^{\mathrm{u}}\leq0|Z_{0}^{\mathrm{u}}=y\right]
=\min\limits_{0\leq y<f^{\mathrm{max}}}\Pr\!\left[Z_{J^{\mathrm{u}}}^{\mathrm{u}}\geq 0|Z_{0}^{\mathrm{u}}=y\right]$, we obtain
\begin{align*}
\max_{0\leq x<g^{\mathrm{max}}}\mathbb{E}\bigg[V\bigg| & X_{0}=\frac{A}{2}-x\bigg]  
\,\leq\;
\frac{\max_{0\leq x<g^{\mathrm{max}}}\mathbb{E}\big[P\big|X_{0}=\frac{A}{2}-x\big]+\max_{0\leq y<f^{\mathrm{max}}}\mathbb{E}\big[J^{\mathrm{u}}\big|Z_{0}^{\mathrm{u}}=y\big]}{\min_{0\leq y<f^{\mathrm{max}}}\Pr\!\big[Z_{J^{\mathrm{u}}}^{\mathrm{u}}\geq\frac{A}{2}\big|Z_{0}^{\mathrm{u}}=y\big]}.
\end{align*}
Similarly, it can be shown that
\allowdisplaybreaks
\begin{align*}
\min_{0\leq x<g^{\mathrm{max}}}\mathbb{E}\!\bigg[V\bigg| & X_{0}=\frac{A}{2}-x\bigg]  
\,\geq\;
\frac{\min_{0\leq x<g^{\mathrm{max}}}\mathbb{E}\big[P\big|X_{0}=\frac{A}{2}-x\big]+\min_{0\leq y<f^{\mathrm{max}}}\mathbb{E}\big[J^{\mathrm{u}}\big|Z_{0}^{\mathrm{u}}=y\big]}{\max_{0\leq y<f^{\mathrm{max}}}\Pr\big[Z_{J^{\mathrm{u}}}^{\mathrm{u}}\geq\frac{A}{2}\big|Z_{0}^{\mathrm{u}}=y\big]}.
\end{align*}
Bounds on $\mathbb{E}\big[P\big|X_{0}=\frac{A}{2}-x\big]$ can be derived in an analogous manner, i.e.,
by conditioning on hitting first the bottom of the buffer or leaving the lower half of the buffer:
\allowdisplaybreaks
\begin{flalign*}
\mathbb{E}\!\left[P\bigg|X_{0}=\frac{A}{2}-x\right] = \mathbb{E}\big[J^{\mathrm{l}}\big|Z_{0}^{\mathrm{l}}=-x\big] 
+\Pr\!\left[Z_{J^{\mathrm{l}}}^{\mathrm{l}}\leq-\frac{A}{2}\bigg|Z_{0}^{\mathrm{l}}=-x\right]\frac{\mathbb{E}\left[K^{\mathrm{l}}\big|Z_{0}^{\mathrm{l}}=0\right]}{\Pr\!\left[Z_{K^{\mathrm{l}}}^{\mathrm{l}}\geq\frac{A}{2}\big|Z_{0}^{\mathrm{l}}=0\right]}.
\end{flalign*}
Using Theorem 1, we can estimate the above quantities as follows:
\begin{align*}
\mathbb{E}\!\left[P\bigg|X_{0}=\frac{A}{2}-x\right] 
& = \Theta\big(\mathbb{E}\big[K^{\mathrm{l}}\big|Z_{0}^{\mathrm{l}}=0\big]\big)
  = \Theta\left(\frac{\mathbb{E}\!\left[Z_{K^{\mathrm{l}}}^{\mathrm{l}}\big|Z_{0}^{\mathrm{l}}=0\right]}{\mathbb{E}\big[W_{1}^{\mathrm{d}}(\hat{X}_{k}=0)\big]}\right)
  = \Theta(\delta^{-1}_{\mathrm{d}}),\\
\Pr\!\left[Z_{\Ku}^{\mathrm{u}}\leq-\frac{A}{2}\bigg|Z_{0}^{\mathrm{u}} = 0\right]
& = 1-\Pr\!\left[Z_{\Ku}^{\mathrm{u}}\geq0|Z_{0}^{\mathrm{u}}=0\right]=\Theta(1),\\
\mathbb{E}\left[J^{\mathrm{u}}|Z_{0}^{\mathrm{u}}=y\right]
& = \Theta(1),\\
\Pr\!\left[Z_{J^{\mathrm{u}}}^{\mathrm{u}}\geq\frac{A}{2}\bigg|Z_{0}^{\mathrm{u}} = y\right]
& =\Theta\!\left(\exp\!\left(-\frac{A}{2}r_{\mathrm{d}}^{*}(\hat{X}_{k}=1)\right)\right),\\
\mathbb{E}\left[\Ku|Z_{0}^{\mathrm{u}}=0\right]
& =\Theta(A),
\end{align*}
where $\delta_{\mathrm{d}}=\mathbb{E}[W_{1}^{\mathrm{d}}(\hat{X}_{k}=0)]$
is the drift in the lower half of the buffer.
Inserting these estimates into (\ref{eq:2.3}) leads to
\begin{equation}\label{eq:Ey_est}
\mathbb{E}\left[Y\right]=\Theta\!\left(\delta_{\mathrm{d}}^{-1}\exp\!\left(\frac{A}{2}r_{\mathrm{d}}^{*}(\hat{X}_{k}=1)\right)\right).
\end{equation}
Lemma \ref{lemma3} implies 
\allowdisplaybreaks
\begin{align}
r_{\mathrm{d}}^{*}(\hat{X}_{k}=1) & =\frac{2\delta_{\mathrm{d}}}{\mathrm{var}(W_{k}^{\mathrm{d}}(\hat{X}_{k}=1)|\delta_\mathrm{d}=\delta_\mathrm{e}={0}) + O(|\delta_{\mathrm{e}}|)}  - O(\delta_{\mathrm{d}}^{2})
\nonumber\\
 & =
 \frac{\ln A}{A}
 \frac{2\beta_{1}}{\mathrm{var}(W_{k}^{\mathrm{d}}(\hat{X}_{k}=1)|\delta_\mathrm{d}=\delta_\mathrm{e}={0})+O(|\delta_{\mathrm{e}}|)}-O\!\left(\!\left(\frac{\ln A}{A}\right)^{\!2}\right),\label{eq:rdx1}
\end{align}
where $\delta_\mathrm{d}=\frac{\beta_{1}\ln A}{A}$ and
the term $O(|\delta_{\mathrm{e}}|)$ accounts for the variation of the second-order statistic of the Markov chain $W_{k}^{\mathrm{d}}(\hat{X}_{k})$
around $\delta_\mathrm{d}=\delta_\mathrm{e}={0}$.

Combining $p_{\mathrm{FQ}}^{\pi^\mathrm{o}} = \frac{1}{\mathbb{E}[Y]}$, \eqref{eq:Ey_est}, and \eqref{eq:rdx1}, 
and assuming
$\beta_{1}/2 > \mathrm{var}(W_{k}^{\mathrm{x}}(\hat{X}_{k})|\delta_\mathrm{d}=\delta_\mathrm{e}={0})+O(|\delta_{\mathrm{e}}|)$,
we arrive at
\allowdisplaybreaks
\begin{flalign*}
p_{\mathrm{FQ}}^{\pi^\mathrm{o}}
& 
  = O\!\left(\delta_{\mathrm{d}} \exp\!\Big(-\frac{A}{2}r_{\mathrm{d}}^{*}(\hat{X}_{k}=1)\Big)\right)\\
& = O\!\left(\frac{\ln A}{A} \exp\!\left[\ln A\frac{-\beta_{1}}{\mathrm{var}(W_{k}^{\mathrm{d}}(\hat{X}_{k}=1)|\delta_\mathrm{d}=\delta_\mathrm{e}={0})+O(|\delta_{\mathrm{e}}|)}+O\!\left(\frac{\ln^2\!A}{A}\right)\right]\right)\\
 & =o\!\left(\frac{1}{A^{2}}\right).
\end{flalign*}

A similar derivation can be used to show that the probability for an empty energy buffer scales as
$p_{\mathrm{EB}}^{\pi^\mathrm{o}}=o(1/B^{2})$. Here, 
the decremental drift is chosen as $\delta_\mathrm{e}=\frac{\beta_{2}\ln B}{B}$ 
and we need to choose 
with $\beta_{2}/2> \mathrm{var}(W_{k}^{\mathrm{e}}(\hat{E}_{k})|\delta_\mathrm{d}=\delta_\mathrm{e}={0})+O(|\delta_{\mathrm{d}}|)$.
We note that the estimates for $p_{\mathrm{FQ}}^{\pi^\mathrm{o}}$ and $p_{\mathrm{EB}}^{\pi^\mathrm{o}}$ 
are tight if the renewal epochs of the Markov chains $\tilde{X}_{k}$ and $\tilde{E}_{k}$ are sufficiently small
compared to $A$ and $B$, respectively (see \cite{Tse-JSAC} and references therein for further details).

We next analyze the convergence of the average distortion to the optimal value. The average distortion during normal operation 
as defined in \eqref{eq:Dop_def} can be expressed via the Taylor expansion
\begin{align*}
D^{\mathrm{op}} = &\;
 D_{\mathrm{T}}(0,0) +
\frac{\partial D_{\mathrm{T}}(0,0)}{\partial \delta_{\mathrm{d}}}\frac{\beta_{1}\ln A}{A} (2q_\mathrm{d}-1)
+
\frac{\partial D_{\mathrm{T}}(0,0)}{\partial \delta_{\mathrm{e}}}\frac{\beta_{2}\ln B}{B} (2q_\mathrm{e}-1) 
\\ & 
+\frac{\partial^{2}D_{\mathrm{T}}(0,0)}{2\partial \delta_{\mathrm{d}}^2}\left(\frac{\beta_{1}\ln A}{A}\right)^{\!2}
+\frac{\partial^{2}D_{\mathrm{T}}(0,0)}{2\partial \delta_{\mathrm{e}}^2}\left(\frac{\beta_{2}\ln B}{B}\right)^{\!2}
+ O\!\left(\!\left(\frac{\beta_{1}\ln A}{A}\right)^{\!3}+\left(\frac{\beta_{2}\ln B}{B}\right)^{\!3}\right),
\end{align*}
\allowdisplaybreaks
where $q_\mathrm{d}=\Pr(\hat{X}_{k}=1)$ and $q_\mathrm{e}=\Pr(\hat{E}_{k}=1)$. 

We next show that the scaling behavior of the first-order terms in this expansion is $o(1/A^{2})$ and $o(1/B^{2})$, respectively.
Since the data buffer is finite, the difference between input and output in steady state equals zero. 
Mathematically,
\begin{equation}\label{eq:dbuff_zeromean}
\lim_{n\rightarrow\infty}\frac{1}{n}\left[ \sum_{k=0}^{n}W_{k}^{\mathrm{d}}+L_{n}^{\mathrm{d}}-U_{n}^{\mathrm{d}} \right]= 0,
\end{equation}
where $W_{k}^{\mathrm{d}}=f(D_{k},E_{\mathrm{s},k},Q_{k})-g(H_{k},E_{\mathrm{t},k})$
is the data buffer increment process
(neglecting boundary effects), 
$L_{n}^{\mathrm{d}}$ is the cumulative number of bits up to time $n$ that have been padded with zeros due to underflow, 
and $U_{n}^{\mathrm{d}}$ is the cumulative number of bits up to time $n$ that have been lost due to overflow.
According to the strong law of large numbers for renewal-reward processes \cite{BOR76}, 
\[
\lim_{n\rightarrow\infty}\frac{1}{n}\sum_{k=0}^{n}W_{k}^{\mathrm{d}}
= \mathbb{E}\!\left[W_{1}^{\mathrm{d}}\right]=\frac{\beta_{1}\ln A}{A} (2q_\mathrm{d}-1)
.
\]
As to the boundary effects, there is a one-to-one correspondence between the times at which $U_{n}^{\mathrm{d}}$ increases and
the times that the buffer is full. Furthermore, this increase is bounded by $f^{\mathrm{max}}$, so that
\begin{gather*}
\lim_{n\rightarrow\infty}\frac{U_{n}^{\mathrm{d}}}{n}\leq f^{\mathrm{max}}p_{\mathrm{FQ}}^{\pi}=o(1/A^{2})
\end{gather*}
Due to the symmetry of the problem, we similarly obtain $\lim_{n\rightarrow\infty}L_{n}^{\mathrm{d}}/n=o(1/A^{2})$.
Hence, \eqref{eq:dbuff_zeromean} implies
$\frac{\beta_{1}\ln A}{A} (2q_\mathrm{d}-1) = o(1/A^{2})$.
The same line of arguments can be used to show 
$\frac{\beta_{2}\ln B}{B} (2q_\mathrm{e}-1) = o(1/B^{2})$.
By combining the above intermediate results, it follows that the average distortion during normal operation
reads $D^{\mathrm{op}}=D_{\mathrm{T}}(0,0)+O\left(\left(\frac{\beta_{1}\ln A}{A}\right)^{2}\right)+O\left(\left(\frac{\beta_{2}\ln B}{B}\right)^{2}\right).$
Inserting this expression for $D^{\mathrm{op}}$ 
together with $p_{\mathrm{FQ}}^{\pi^\mathrm{o}}=o(1/A^{2})$ and $p_{\mathrm{EB}}^{\pi^\mathrm{o}}=o(1/B^{2})$ 
into the upper bound (\ref{eq:av_dist})
of the average distortion finally confirms that $\bar{D}^{\pi^\mathrm{o}}$
approaches $\bar{D}^{*}$ as $O((\ln A/A)^{2})+O((\ln B/B)^{2})$.
Proposition 3 is thus proved.

\bibliographystyle{ieeetr}
\bibliography{IEEEabrv,SPLrefs,proofs_appendix}

\end{document}